\documentclass{PoS}
\usepackage{amsmath,bm}
\newcommand{\diff}{\mathrm{d}}

\newcommand{\msbot}{m_{\widetilde b}}
\newcommand{\sbot}{\widetilde b}
	
\newcommand{\avb}[1]{\big\langle #1 \big\rangle}	
\newcommand{\eq}{\mathrm{eq}}
\newcommand{\s}[1]{\widetilde #1}

\newcommand{\Ysbot}{Y_{\sbot}}

\def\beq{\begin{equation}}
\def\eeq{\end{equation}}

\title{
Conversion-driven freeze-out: Dark matter genesis beyond the WIMP paradigm
}

\ShortTitle{Conversion-driven freeze-out}

\author{Mathias Garny,$^{\!a\,}$ \speaker{Jan Heisig},$^{\!b\,}$ Marco Hufnagel,$^{\!c\,}$ Benedikt L{\"u}lf$^{\,d}$ and Stefan Vogl$^{\,e}$
       \\
       \\
      \llap{$^{a}$}
        Technische Universit\"at M\"unchen, James-Franck-Str. 1, 85748 Garching,
Germany\\
        \llap{$^{b}$}
        Centre for Cosmology, Particle Physics and Phenomenology (CP3), Universit\'e catholique de Louvain, Chemin du Cyclotron 2, B-1348 Louvain-la-Neuve, Belgium\\
          \llap{$^{c}$}
          DESY, Notkestra\ss e 85, D-22607 Hamburg, Germany \\
        \llap{$^{d}$} Institute for Theoretical Particle Physics and Cosmology,
RWTH Aachen University, Sommerfeldstr. 16, 52056 Aachen, Germany\\
\llap{$^{e}$}Max-Planck-Institut f\"ur Kernphysik, Saupfercheckweg 1, 69117 Heidelberg,
Germany \vspace{0.6ex}\\ 
	{\footnotesize E-mail: \email{mathias.garny@tum.de}, 
	\email{jan.heisig@uclouvain.be}, 
	\email{marco.hufnagel@desy.de}, 
	\email{benedikt.luelf@gmx.de}, 
	\email{stefan.vogl@mpi-hd.mpg.de} }
	}

\abstract{
We consider dark matter (DM) with very weak couplings to the standard model (SM), such that its
self-annihilation cross section is much smaller than the canonical one, 
$\langle\sigma v\rangle_{\chi\chi} \ll 10^{-26}\mathrm{cm}^3/\mathrm{s}$. 
In this case DM self-annihilation is 
negligible for the dynamics of freeze-out
and DM dilution is solely driven by efficient annihilation of heavier accompanying dark sector particles
provided that DM maintains chemical equilibrium with the dark sector.
This chemical equilibrium is established by conversion processes which require much smaller couplings
to be efficient than annihilation.
The chemical decoupling of DM from the SM can either be initiated by the freeze-out 
of annihilation, resembling a co-annihilation scenario, or of conversion processes, leading to
the scenario of conversion-driven freeze-out. We focus on the latter and discuss its distinct phenomenology.
}

\FullConference{Corfu Summer Institute 2018 "School and Workshops on Elementary Particle Physics and Gravity"\\
		(CORFU2018)\\
		31 August - 28 September, 2018\\
		Corfu, Greece}

\begin{document}

\section{Introduction}\label{sec::intro}

A wide range of cosmological and astrophysical observations require the extension of the standard model (SM) 
of particle physics by a (cold) dark matter (DM) candidate, see \emph{e.g.}~\cite{Bertone:2004pz}
for a review on its evidence.
Its abundance in the Universe today can be elegantly explained by a thermal relic from early Universe dynamics.
Among such scenarios, thermal freeze-out is one of the simplest and most attractive ones.
In this case DM is assumed to be in thermal contact in the very early Universe
and decouples once the temperature drops significantly below the DM mass. Appealing features of this mechanism are that the resulting abundance is independent of the not very-well constrained physics prior to freeze-out (\emph{e.g.}~inflation and reheating) and that DM is guaranteed to be sufficiently cold to allow for successful structure formation. 
In the simplest freeze-out scenario the break-down of DM self-annihilation initiates the chemical decoupling
of DM from the SM\@. In this case, imposing the relic abundance to match the observed value, $\Omega_\mathrm{CDM} h^2\simeq 0.12$, requires the respective thermally averaged annihilation cross section to be of the order of 
$\langle\sigma v\rangle_{\chi\chi} \sim g_\chi^4/m_\chi^2 \sim 10^{-9} \,\mathrm{GeV}^{-2}$.
This relation singles out an allowed band in the plane spanned by the DM mass, $m_\chi$, and 
the characteristic DM-SM coupling strength, $g_\chi$. It contains the weakly interacting massive particle (WIMP)
as a special case, $m_\chi\sim v_\mathrm{EW}$, $g_\chi\sim g_\mathrm{EW}$.

A variety of ongoing experiments probe large portions of this band.  However, most searches are subject to a large model-dependence. 
For models with dominant $s$-wave annihilation the most direct test of the self-annihilating nature of DM is provided by indirect detection experiments. 
Masses below a few GeV are excluded by limits from the non-observation of effects of energy injection during recombination in the 
cosmic microwave background~\cite{Ade:2015xua}. These limits are rather independent of the actual annihilation process and do not contain uncertainties from the DM density profile. For larger masses limits from Fermi-LAT dwarfs~\cite{Drlica-Wagner:2015xua} and eventually from cosmic-ray antiprotons~\cite{Cuoco:2017iax} constrain the above scenario,
reaching up to $m_\chi\simeq500\,\mathrm{GeV}$ for non-leptonic annihilation channels.

These results and a variety of additional experimental constraints narrow down the allowed parameter space of 
the simplest models of thermal relics. Hence it is interesting to consider alternative thermal scenarios that do not just 
evade these bounds but could potentially point to novel and unexplored signatures of DM\@.
In this article we consider scenarios where the self-annihilation of
DM is not responsible for the dynamics of freeze-out. In fact, we assume it to be negligible
$\langle\sigma v\rangle_{\chi\chi}
\ll 10^{-9} \,\mathrm{GeV}^{-2}$. This can be realized once a second, slightly\footnote{For a viable exception see \emph{e.g.}~\cite{DAgnolo:2018wcn}.} heavier 
particle is contained in the dark sector and drives the dilution of the dark sector abundances
as it can be the case in the co-annihilation scenario~\cite{Edsjo:1997bg}. In this scenario
conversion processes within the dark sector are assumed to be efficient thus maintaining chemical equilibrium 
among dark sector particles. Consequently, chemical decoupling is usually expected to be initiated by
the break-down of annihilations. However, chemical decoupling can 
also be initiated by the break-down of conversion processes leading to conversion-driven freeze-out~\cite{Garny:2017rxs} (or co-scattering~\cite{DAgnolo:2017dbv}). This latter possibility can be realized for very small DM-SM couplings. 

In the Sec.~\ref{sec:2} we review the general features of conversion-driven freeze-out. A realization within a simple 
$t$-channel mediator DM model, its numerical solutions and constraints are discussed in Sec.~\ref{sec:3}. We conclude in
Sec.~\ref{sec::sum}.

\section{Conversion versus annihilations}\label{sec:2}

A commonly made assumption in co-annihilation scenarios is that conversion processes 
between DM and the slightly heavier co-annihilating partner(s) are thoroughly efficient during freeze-out.
In this case, relative chemical equilibrium is maintained,  
$n_{\chi_i}/n_{\chi_i}^{\eq}=n_{\chi_j}/n_{\chi_j}^{\eq}$, 
and annihilations can be described by an effective, 
thermally averaged cross section\,\cite{Edsjo:1997bg}
\begin{equation}
\label{eq:effsigmnav}
\langle\sigma v\rangle_\mathrm{eff} = \sum_{i,j}\langle\sigma v\rangle_{ij}\frac{n_{\chi_i}^\eq}{n_\chi^\eq}\frac{n_{\chi_j}^\eq}{n_\chi^\eq}\,,
\end{equation}
where $n_\chi^\eq=\sum_i n_{\chi_i}^\eq$ and $i,j$ runs over all dark sector particles. 

For a large hierarchy in the couplings to the SM between DM and the co-annihilating partner(s), 
$\langle\sigma v\rangle_{ij}$ could be negligible for all channels containing DM in the initial state.
Dark matter dilution is then entirely driven by annihilations of heavier states.

The rate of conversion processes, which contain light SM degrees of freedom in the initial state, is expected to be
boosted relative to the rate of dark sector annihilations for $T < m_{\chi}$:
\begin{equation}
\label{eq:rats}
\frac{\Gamma_\mathrm{con}}{\Gamma_\mathrm{ann}}
\sim \frac{n^\eq_a}{n^\eq_{\chi}}\frac{\langle\sigma v\rangle_{\chi_i a\to \chi_j b}}{\langle\sigma v\rangle_\text{eff}}
\sim \mathrm{e}^{m_{\chi}/T}  \frac{\langle\sigma v\rangle_{\chi_i a\to \chi_j b}}{\langle\sigma v\rangle_\text{eff}}\,,
\end{equation}
where $a,b$ are SM particles. 
For simplicity, in \eqref{eq:rats} we only consider conversions via a $2\to2$ scattering cross section.  However, inverse decays
or $2\leftrightarrow3$ scattering processes can also be important in general.

There are two regimes regarding the size of the conversion terms and hence the temperature where the chemical decoupling (indicated by the subscript CD) in the dark sector takes place:
\begin{itemize}
\item[(i)] 
$T_\mathrm{CD,\,con} \ll T_\mathrm{CD,\,ann}$: Chemical equilibrium within the dark sector breaks down well after the
chemical decoupling of the dark sector from the SM takes place. This resembles the usual co-annihilation scenario. 
\item[(ii)] 
$T_\mathrm{CD,\,con} \gtrsim T_\mathrm{CD,\,ann}$: Chemical decoupling of DM from the dark sector takes place before or during 
chemical decoupling of the dark sector from the SM\@. This regime is called conversion-driven freeze-out~\cite{Garny:2017rxs} (or co-scattering~\cite{DAgnolo:2017dbv}). 
\end{itemize}
In the latter case $\Gamma_\mathrm{con}$ determines the relic density. 
For a typical freeze-out temperature of $T_\mathrm{CD,\,ann} \,\sim \,m_\chi/25$, and considering  
$\Gamma_\mathrm{con}/\Gamma_\mathrm{ann}\sim 1$ at chemical decoupling, $\langle\sigma v\rangle_{\chi_i a\to \chi_j b}$ could 
be smaller than $\langle\sigma v\rangle_\text{eff}$ by a factor of $\mathrm{e}^{-25}\sim 10^{-11}$. Accordingly, conversion-driven freeze-out is realized for very weak DM couplings to the SM\@.

\section{Numerical example}\label{sec:3}

\subsection{A minimal model}

We consider one of the simplest example models providing a co-annihillation dark sector, namely a simplified 
dark matter model with a $t$-channel mediator:
\begin{equation}
    \mathcal{L}_\mathrm{int} = |D_\mu \tilde q|^2 + \lambda_\chi \tilde q\, \bar{q}\,\frac{1-\gamma_5}{2}\chi +\mathrm{h.c.}\,,
    \label{eq:tchmodel}
\end{equation}
where $q$ is a SM quark field, $\tilde q$ is a scalar partner of the quark (sharing the same gauge quantum numbers) and the DM $\chi$ is a Majorana fermion.
Here $D_\mu$ denotes the covariant derivative, $(1-\gamma_5)/2$ is the left-handed projection operator
and $\lambda_\chi$ is the DM coupling strength.
The model is similar to a limiting case of supersymmetry where $\chi$ is a bino-like neutralino and $\tilde q$
is a right-handed squark. Unlike in the minimal supersymmetric SM, $\lambda_\chi$ is not set by
the SM gauge couplings but considered to be a free parameter of the model.

The model respects a $Z_2$-symmetry under which all SM fields are even
while $\chi$ and $\tilde q$ are odd. An additional Higgs portal coupling of type $\Phi^\dagger \Phi\,\tilde q^\dagger \tilde q$ is
allowed by the symmetries as well. However, in the following we assume it to be negligible for simplicity since it does not affect the qualitative
picture. For a more detailed discussion of the model and its phenomenology see for example~\cite{Garny:2017rxs,Ibarra:2015nca,Garny:2018icg}.
We consider the two cases of a bottom and top partner, $q = b, t$.

\subsection{Evolution of abundances}

For the general case where neither $T_{\mathrm{CD,\,con}} \ll T_{\mathrm{CD,\,ann}}$ nor 
$T_{\mathrm{CD,\,con}} \gg T_{\mathrm{CD,\,ann}}$ is guaranteed  
a set of coupled Boltzmann equations has to be solved. For the minimal model considered here they read
\begin{eqnarray}
		\frac{\diff Y_{\chi }}{\diff x} &= &\frac{1}{ 3 H}\frac{\mbox{d} s}{\mbox{d} x}
		\left[\,\avb{\sigma_{\chi\chi}v}\left(Y_{\chi}^2-Y_{\chi}^{\eq\,2}\right)+\avb{\sigma_{\chi\tilde q} v}\left(Y_{\chi}Y_{\tilde q}-Y_{\chi}^{\eq}Y_{\tilde q}^{\eq}\right)
		 \phantom{\left(Y_{\chi}^2-Y_{\tilde q}^2\frac{Y_{\chi}^{\eq\,2}}{Y_{\tilde q}^{\eq\,2}}\right)} \right. \qquad\nonumber \\
		 &&\left.+\;\frac{\Gamma_{\chi\rightarrow \tilde q}}{s}\left(Y_{\chi}-Y_{\tilde q}\frac{Y_{\chi}^{\eq}}{Y_{\tilde q}^{\eq}}\right) -\frac{\Gamma_{\tilde q} }{s}\left(Y_{\tilde q}-Y_{\chi}\frac{Y_{\tilde q}^{\eq}}{Y_{\chi}^{\eq}}\right)+\avb{\sigma_{\chi\chi\rightarrow \tilde q\tilde q^\dagger}v}\left(Y_{\chi}^2-Y_{\tilde q}^2\frac{Y_{\chi}^{\eq\,2}}{Y_{\tilde q}^{\eq\,2}}\right) \right] \label{eq:BMEchi}\,,\quad\quad\\
		\frac{ \diff Y_{\tilde q } }{\diff x} &= & \frac{1}{ 3 H}\frac{\mbox{d} s}{\mbox{d} x}
		\left[ \,\frac{1}{2}\avb{\sigma_{\tilde q\tilde q^\dagger}v}\left(Y_{\tilde q}^2-Y_{\tilde q}^{\eq\,2}\right)+\avb{\sigma_{\chi \tilde q}v}\left(Y_{\chi}Y_{\tilde q}-Y_{\chi}^{\eq}Y_{\tilde q}^{\eq}\right) 
		\phantom{\left(Y_{\chi}^2-Y_{\sbot}^2\frac{Y_{\chi}^{\eq\,2}}{\Ysbot^{\eq\,2}}\right)} \right. \qquad\nonumber \\
		 &&\left.-\;\frac{\Gamma_{\chi\rightarrow \tilde q}}{s}\left(Y_{\chi}-Y_{\tilde q}\frac{Y_{\chi}^{\eq}}{Y_{\tilde q}^{\eq}}\right) +\frac{\Gamma_{\tilde q}}{s}\left(Y_{\tilde q}-Y_{\chi}\frac{Y_{\tilde q}^{\eq}}{Y_{\chi}^{\eq}}\right)
		 -\avb{\sigma_{\chi\chi\rightarrow \tilde q\tilde q^\dagger}v}\left(Y_{\chi}^2-Y_{\tilde q}^2\frac{Y_{\chi}^{\eq\,2}}{Y_{\tilde q}^{\eq\,2}}\right) \right]\label{eq:BMEsqu}\,,
\end{eqnarray}
where  $Y_i= n_i/s$ is the comoving number density for the species $i$, 
$s$ is the entropy density, $H$ denotes the Hubble parameter while $x=m_\chi/T$ parametrizes the temperature dependence. The first two terms in brackets in each equation
represent annihilations into SM particles, while the last three terms (second lines) represent conversions within the
dark sector. Among these $\Gamma_{\chi\rightarrow \tilde q}$ is the thermally averaged rate for $2\to2$ (or $2\leftrightarrow3$, relevant for $q=t$)
scattering, $\Gamma_{\tilde q}$ is the thermally averaged decay rate while the last term represents pair conversion within the dark sector. The latter can usually be neglected as it is subdominant ($\mathcal{O}(\lambda_\chi^4)$) whenever chemical equilibrium is questionable (and hence $\lambda_\chi$ is small).

\begin{figure}[t]
\centering
\setlength{\unitlength}{1\textwidth}
\begin{picture}(0.96,0.69)
  \put(0.0,-0.014){\includegraphics[width=0.95\textwidth]{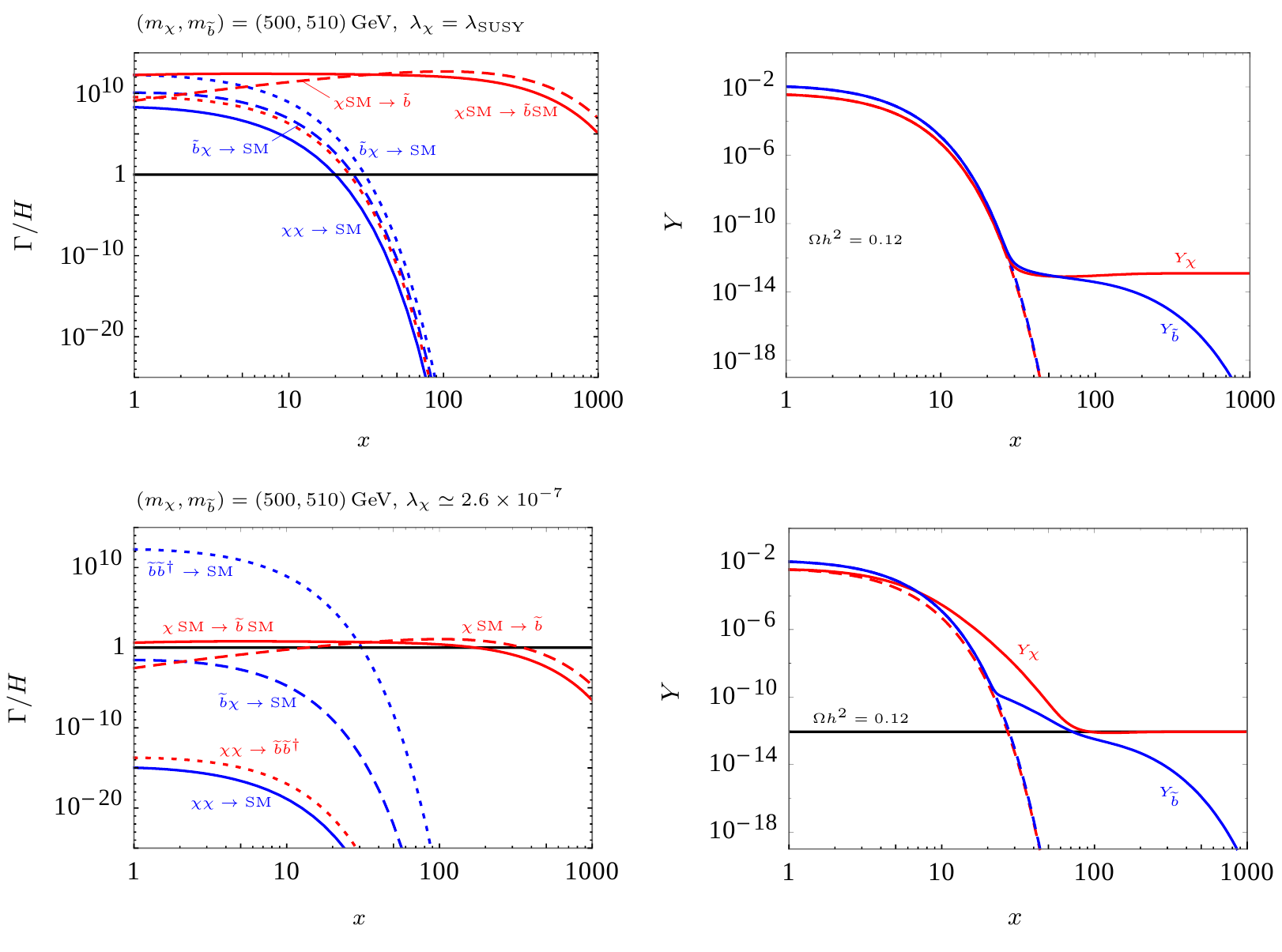}}
\end{picture}
\caption{Left panels: Rates of annihilation (blue curves) and conversion (red curves) terms in the Boltzmann equation relative to
the Hubble rate. Right panels:~Evolution of the resulting abundance (solid curves) of $\s b$ (blue) and $\chi$ (red) as well as the corresponding equilibrium abundances (dashed curves). Both quantities are plotted as a function of the temperature parameter $x=m_\chi/T$.
We consider $m_\chi=500\,$GeV, $\msbot=510\,$GeV as well as $\lambda_\chi=\lambda_\text{SUSY}\simeq 0.17$ (upper panels)
and $\lambda_\chi\approx 2.6\times 10^{-7}$ (lower panels)~\cite{Garny:2017rxs}.
}
\label{fig:ratesY}
\end{figure}

For sizeable couplings $\lambda_\chi$ the conversion rates are much faster than the Hubble rate such that chemical equilibrium in the dark sector is a good assumption. This can be used to reduce 
Eqs.~\eqref{eq:BMEchi} and \eqref{eq:BMEsqu} to a single equation in terms of $Y=\sum_i Y_i$ and the effective annihilation cross section defined in Eq.~\ref{eq:effsigmnav}.\footnote{This is commonly done in
numerical relic density solvers~\cite{Belanger:2018mqt,Ambrogi:2018jqj,Bringmann:2018lay}.} 
This case certainly applies to supersymmetry where $\lambda_\chi=0.17\, (0.33)$ for  
co-annihilation with a right-handed sbottom (stop). The respective relative rates and abundances for the
case of the sbottom are shown in the upper panels of Fig.~\ref{fig:ratesY} for the parameter point with 
$m_\chi=500\,$GeV, $\msbot=510\,$GeV. For this parameter point DM would be under-abundant.

For much smaller couplings, $\lambda_\chi\lesssim 10^{-6}$, conversion processes start to become
inefficient during the freeze-out of $\tilde b$, $\Gamma_\mathrm{con}/H\sim 1$. Therefore the solution of the coupled set of Boltzmann equations is necessary.\footnote{As the elastic scattering cross section $\chi a \to \chi a$ is suppressed by $\lambda_\chi^4$, at first sight the assumption of kinetic equilibrium -- justifying the solution of the \emph{integrated}
Boltzmann equations \eqref{eq:BMEchi} and \eqref{eq:BMEsqu} -- appears questionable as well. However, as shown
explicitly in~\cite{Garny:2017rxs}, the DM momentum distribution does not deviate significantly from
the thermal one and the integrated equations provide a good approximation. Due to the small mass splitting DM approximately 
inherits the momentum distribution of the mediator, which is kept in kinetic equilibrium throughout the freeze-out process
thanks to its strong coupling to the SM\@.}
The evolution of the relative rates and the abundance is shown for 
$\lambda_\chi\approx 2.6\times 10^{-7}$ which is the coupling that provides the right relic density for
the chosen mass point $m_\chi=500\,$GeV, $\msbot=510\,$GeV. In this case $Y_\chi$ already departs from thermal equilibrium
at around $x\sim3$ while a significant depletion of DM occurs up to $x\simeq100$. Both abundances have a non-trivial evolution. Hence, for this scenario the process of freeze-out extends over a large range of $x$ -- in contrast to a typical WIMP\@. 

\begin{figure}[t]
    \centering
    \setlength{\unitlength}{1\linewidth}
        \begin{picture}(0.52,0.273)
\put(0.015,-0.01){\includegraphics[width=0.475\textwidth]{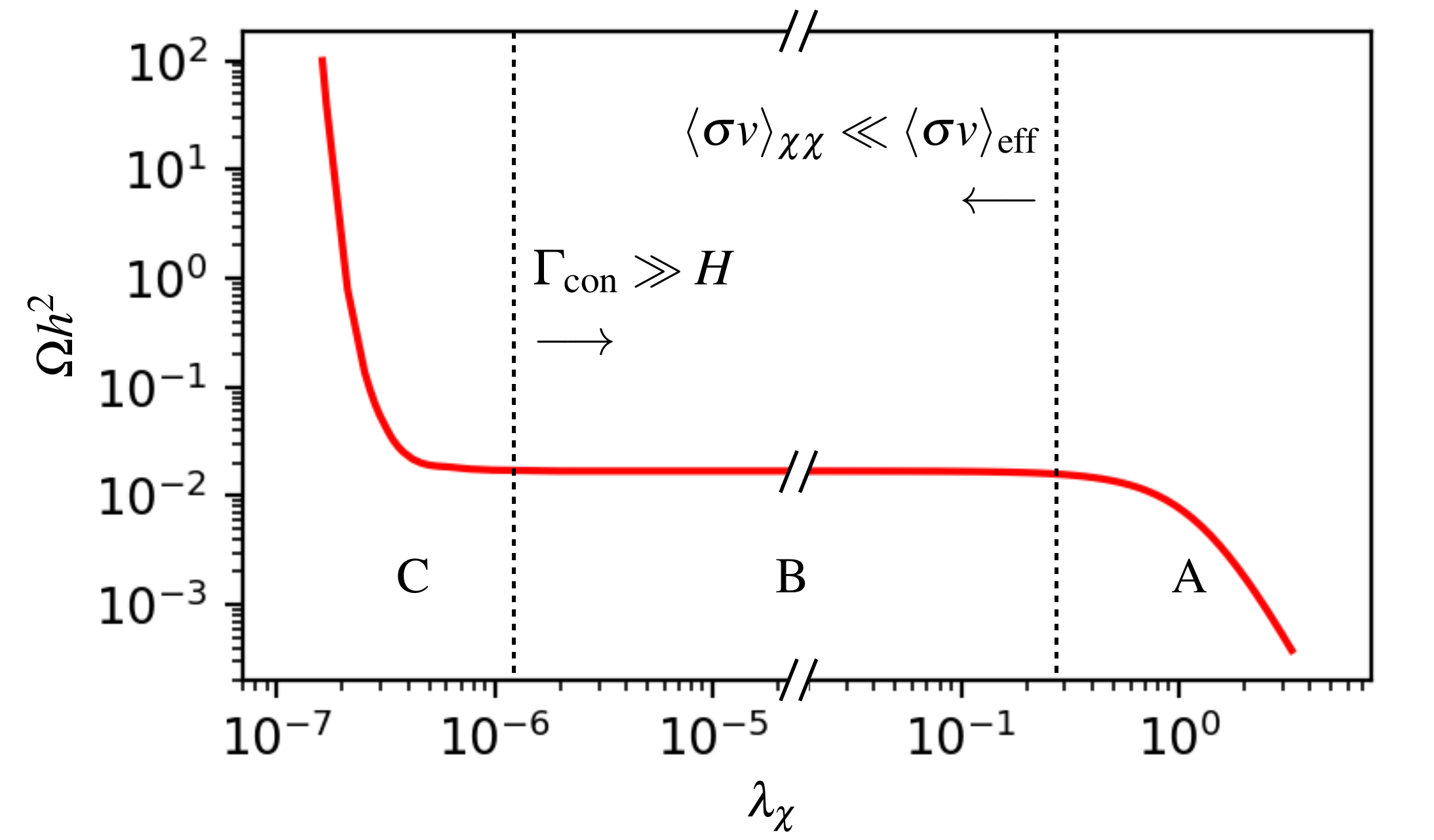}}
        \end{picture}
    \caption{Relic density as a function of the DM coupling $\lambda_\chi$ for 
    $m_\chi=500\,$GeV, $\msbot=510\,$GeV. The regions A, B and C denote the characteristic 
    phenomenological regimes, see text for details.
    }
    \label{fig:rellam}
\end{figure}

Figure~\ref{fig:rellam} shows the resulting relic density as a function of the coupling over the range from $10^{-7}$ to $\mathcal{O}(1)$. 
In the regime of $\mathcal{O}(1)$ couplings $\chi$ self-annihilation or $\chi$-${\tilde b}$ co-annihilation 
are relevant and hence introduce a dependence on $\lambda_\chi$ (region A).
For smaller couplings mediator pair-annihilations dominate the annihilation while $\lambda_\chi$ is still 
large enough to maintain chemical equilibrium in the dark sector, $\Gamma\gg H$ (region B). In this region
the relic density is independent of $\lambda_\chi$. Finally, 
for $\lambda_\chi\lesssim 10^{-6}$ we enter the conversion-driven regime (region C), introducing 
a $\lambda_\chi$-dependence through the conversion rates.

\subsection{Viable parameter range}

\begin{figure}[t]
    \centering
    \setlength{\unitlength}{1\linewidth}
        \begin{picture}(1,0.44)
\put(0.015,-0.02){\includegraphics[width=0.47\textwidth]{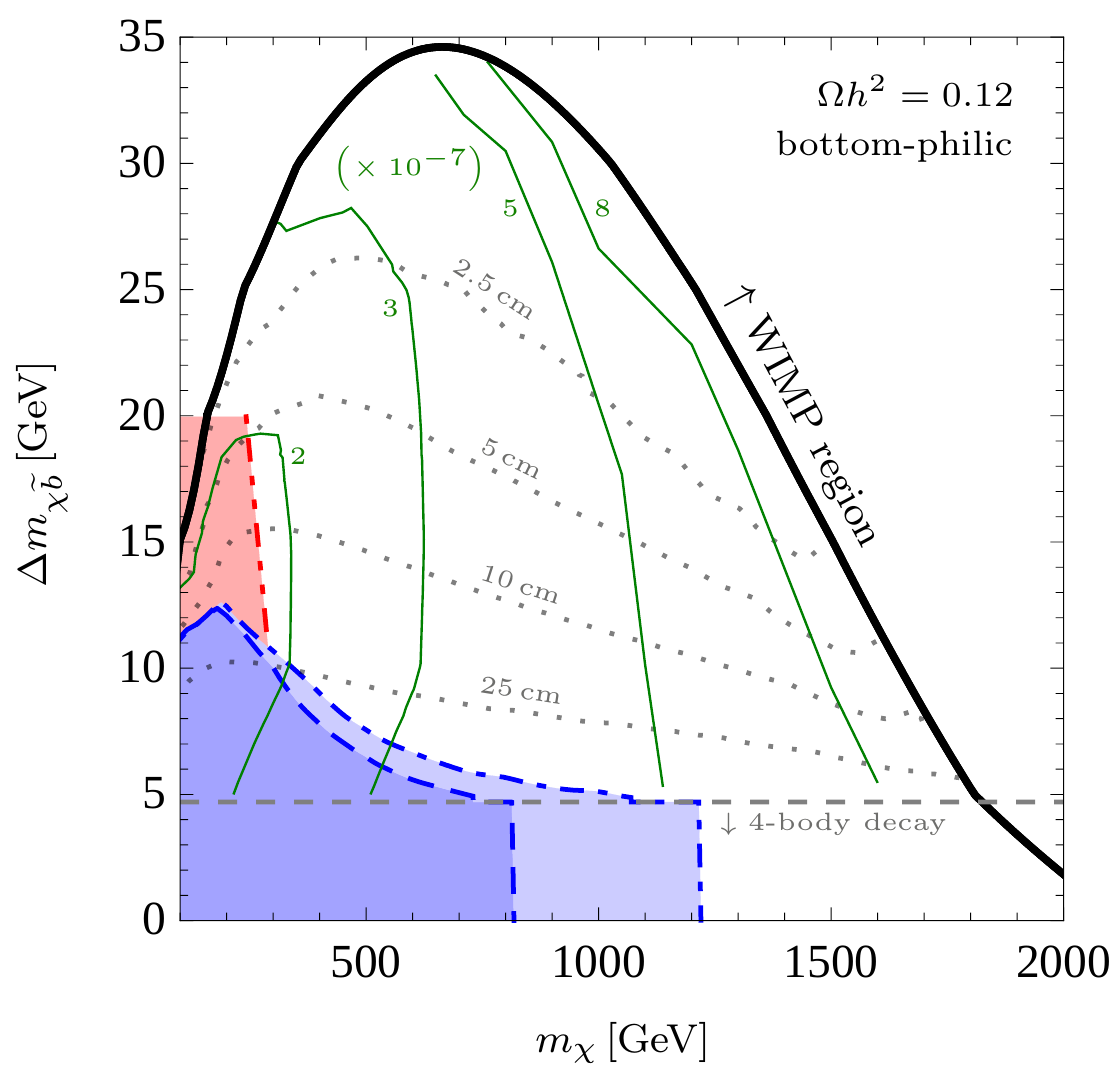}}
\put(0.505,-0.02){\includegraphics[width=0.4788\textwidth]{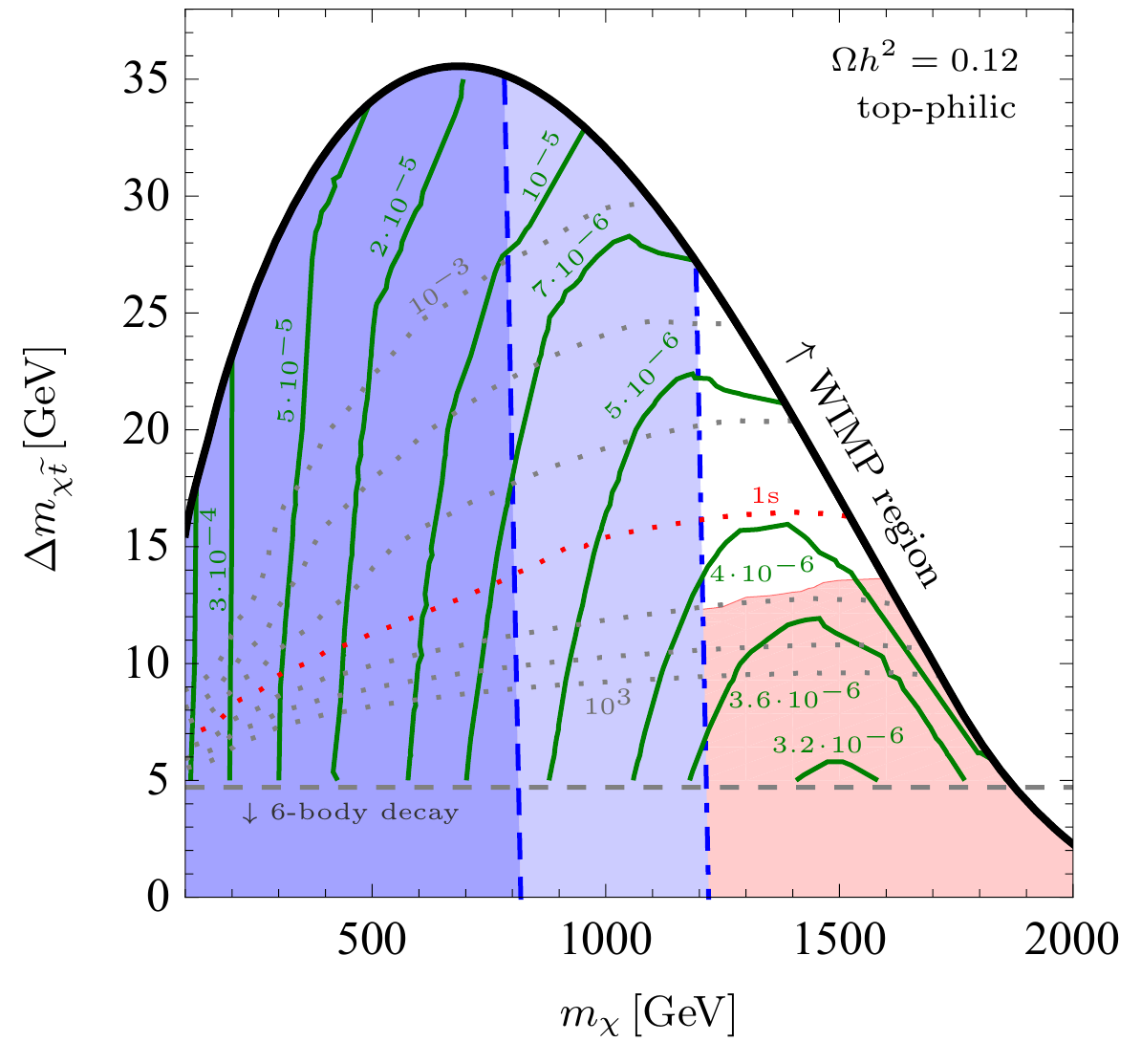}}
        \end{picture}
    \caption{
Cosmologically viable parameter space ($\Omega h^2=0.12$) in the region
providing conversion-driven freeze-out (below black thick curve) in the bottom- (left panel) and top-philic (right panel) model~\cite{Garny:2017rxs,Garny:2018icg}.
We show the contours of constant $\lambda_\chi$ in green ($\times 10^{-7}$ in the left panel).
Contours of constant mediator decay length (left panel) and lifetimes (right panel) are shown as gray dotted curves.
The latter is shown in the range $[10^{-3};\,10^{3}]$s in steps of an order of magnitude (the curve for 1\,s is highlighted in red). 
We show 95\% C.L.~exclusion regions from $R$-hadron searches at the 8 and 13\,TeV LHC in dark and light blue, respectively. 
The red shaded region bordered by the red dot-dot-dashed curve (left panel) denotes the constraint from mono-jet searches at the 13\,TeV LHC\@. The light red shaded region (right panel) indicates constraints from BBN\@. Below the horizontal gray dashed line ($\sim 5\,$GeV) the 2- (left panel) and 4-body decay (right panel) is kinematically forbidden rendering the 4- and 6-body decay, respectively, to be dominant.
    }
    \label{fig:nonCEcont}
\end{figure}

Conversion-driven freeze-out opens up a cosmologically viable region in parameter space 
with small mass splittings between DM and the mediator that yields $\Omega h^2 \simeq 0.12$.
If chemical equilibrium was maintained within the dark sector, efficient co-annihilations would lead to $\Omega h^2 < 0.12$ in that case. 
The respective region in parameter space is shown in Fig.~\ref{fig:nonCEcont} below the thick black curve. 
In fact, the thick black curve separates the regions A (above) and C (below) 
defined in the last paragraph, while region B resides on the curve.
Here we consider the case of the bottom- and top-philic DM model in the left and right panel, respectively. 

The green thin curves in Fig.~\ref{fig:nonCEcont} indicate contours of constant coupling strength $\lambda_\chi$ providing
$\Omega h^2 = 0.12$. The coupling varies between $10^{-7}$ and $10^{-6}$ in the
bottom-philic model and $3\times 10^{-6}$ and $10^{-3}$ for the top-philic model. The reason for the
larger coupling and the larger range spanned in the latter case is due to the fact that all $2\to2$ conversion
processes contain at least one heavy SM particle, namely a $W$-boson or a top-quark, in the initial or final 
state. This leads to an additional Boltzmann suppression (either directly through the $W$ or $t$ abundance
or through phase-space suppression, if particles are in the initial or final state, respectively) 
which is particularly strong for small DM masses. In this regime $2\leftrightarrow3$ processes become dominant.

Figure~\ref{fig:nonCEcont} also displays the resulting lifetime (or decay length) of the mediator
as (gray) dotted curves. The lifetime is drastically different for the two considered models. 
For the bottom-philic model, in a large portion of the allowed parameters space providing 
conversion-driven freeze-out the 2-body decay $\tilde b \to \chi b$ is open and hence the 
decay is only suppressed by the smallness of the coupling $\lambda_\chi$. In fact, for 
the bottom-philic model the decay rate and the $2\to2$ scattering rate during freeze-out are
of the same order (\emph{cf.}~left panels of Fig.~\ref{fig:ratesY}). The resulting decay length
is of the order of a few to a few tenth of a cm. This range provides interesting signatures at colliders,
see Sec.~\ref{sec:sig}.

The situation is vastly different in the top-philic model. In the allowed conversion-driven freeze-out
parameter region the 2-body decay is forbidden and the decay of the mediator proceeds
solely through the 4-body decay. Due to this extra suppression the decay length is always large
compared to the detector-size of colliders. The long lifetime could even be in conflict with BBN for $\tau\gtrsim1\,$sec.
The respective bound utilizing the results from~\cite{Jedamzik:2006xz} is shown as the red shaded region in the
right panel of Fig.~\ref{fig:nonCEcont}. Furthermore, compared to the rates of the leading $2\to2$ or $2\leftrightarrow3$ 
conversion processes the decay rate is negligible during freeze-out.

\subsection{Distinct signature}\label{sec:sig}

Due to the small couplings and the $\lambda_\chi^4$ suppression of the annihilation cross section $\chi\chi\to \text{SM} $ or 
elastic DM-nucleon scattering $\chi \text{SM}\to\chi \text{SM} $ the allowed conversion-driven freeze-out region is not
challenged by indirect or direct detection searches. However, the sizeable coupling required to efficiently annihilate away the
mediator leads to significant interactions of the mediator with the SM and hence 
supports the possibility of searching for the mediator. In our example of a strongly coupled mediator its production 
cross section at the LHC is sizeable. On top of that the mediator decay length  for conversion-driven freeze-out with mediator mass in the GeV--TeV range  is macroscopic. This leads to long-lived particles  which constitute a very distinct signature at the LHC (for a recent, comprehensive account on long-lived particles at the 
LHC see~\cite{Alimena:2019zri}). 

\begin{figure}[t]
    \centering
    \setlength{\unitlength}{1\linewidth}
        \begin{picture}(0.56,0.32)
\put(0.015,-0.02){\includegraphics[width=0.47\textwidth]{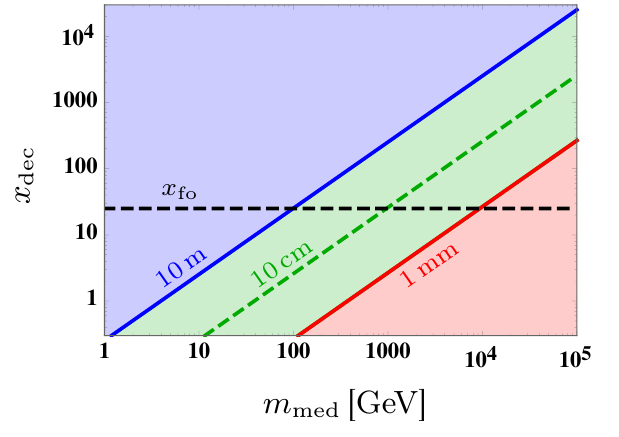}}
        \end{picture}
    \caption{Regions of decay-length in the plane spanned by the mediator mass and the characteristic temperature parameter of its decay, $x_\text{dec}$.
    The blue, green and red regions, respectively, indicate decays to typically take place outside the detector, within the sensitive 
    detector volume or promptly. The dashed black horizontal line indicate a typical value for the temperature parameter
    at freeze-out, $x_\text{fo}=25$.
    }
    \label{fig:xdec}
\end{figure}

Defining $x_\text{dec}$ to be the temperature parameter where the decay is on the edge of being efficient, 
$\Gamma_\mathrm{dec}\sim H(x_\text{dec})$,
and assuming freeze-out to take place in 
the radiation dominated era with the number of relativistic degrees of freedom in the SM being $g_*\simeq100$,
we can derive
\begin{equation}
\label{eq:ctau}
c\tau\sim H^{-1}(x_\text{dec}) \simeq10\,\mathrm{cm} \, \left(\frac{1\,\mathrm{TeV}}{m_\text{med}}\right)^2 \left(\frac{x_\text{dec}}{25}\right)^2\,.
\end{equation}
This general relation between the mass of the mediator
and its decay-length, relevant for the LHC, 
is illustrated in Fig.~\ref{fig:xdec}. It highlights the domains of prompt decays (red shaded),
decays within the detector (green shaded) and detector-stable mediators (blue shaded)
in the plane spanned by the mediator mass and $x_\text{dec}$. 
For the conversion-driven freeze-out scenario considered here
$\Gamma_\text{con}/H\sim 1$ around freeze-out, $x_\text{fo}\sim25$, and hence $x_\text{dec}\sim x_\text{fo}$ or 
$x_\text{dec}\gg x_\text{fo}$, depending on whether the decay provides a sizable contribution to the conversion or not.\footnote{%
To be more precise, in the former case $\Gamma_\text{dec}/H\sim 1$ should rather hold for the rate of the \emph{inverse} decay
which behaves like $\Gamma_\text{dec,\,inv}=n^\eq_{\tilde q}/n^\eq_{\chi} \,\Gamma_\text{dec} \sim \mathrm{e}^{-\Delta m /T}\,
\Gamma_\text{dec}$. In contrast to the decay which becomes more and more efficient with decreasing temperature
($\Gamma_\text{dec}/H\propto T^{-2}$ in the non-relativistic regime and in radiation domination), the inverse decay experiences an additional exponential suppression when $T\lesssim\Delta m$.
It hence reaches its maximum before this suppression becomes sizeable. Therefore, by construction, 
in scenarios where the inverse decay is a relevant contribution to conversion, chemical decoupling takes places at
$T\sim\Delta m$ where the size of the decay and inverse decay rate is of the same order of magnitude.
Nevertheless, we emphasize that, due to a number of neglected $\mathcal O(1)$-factors, eq.~\eqref{eq:ctau} only provides a rough order-of-magnitude estimate.}
In our case the bottom- and top-philic model falls in the first and second class, respectively. 
For the mass range explorable at the current and future updates of the LHC the two scenarios predict lifetimes leading
to decays inside the tracker and detector-stable mediators, respectively. Both signatures provide
promising prospects. 

The (reinterpreted) limits from the 8 and 13 TeV LHC searches for heavy stable charged particles~\cite{Khachatryan:2015lla,CMS-PAS-EXO-16-036}
are shown as the dark and light blue shaded regions in Fig.~\ref{fig:nonCEcont}. As the limits
are suppressed due to the small fraction of mediators in the bottom-philic model that pass
the whole detector (given its small decay length) these limits are much weaker in this case. 
In contrast the limits exclude a significant part of the allowed parameter space for the
top-philic model leaving a small part of the parameter space between around 1.2 and 1.6\,TeV
only which can be probed entirely by 300\,fb$^{-1}$ at the 13 or 14\,TeV LHC~\cite{Garny:2018icg}. 

Another existing search probing a small part of the allowed parameter region of the bottom-philic model
is the mono-jet search~\cite{Aaboud:2016tnv} that solely relies on the recoil against the missing energy caused by DM, not
exploiting the macroscopic decay length of the mediator. The corresponding limit is shown as
the red shaded region in the left panel of Fig.~\ref{fig:nonCEcont}. A dedicated search for 
disappearing $R$-hadron tracks or displaced jets (kinked tracks) is expected to greatly 
strengthen the constraints on the model covering large parts of the yet unexplored region. Current
searches \emph{e.g.}~for disappearing tracks from long-lived charginos are not directly applicable 
to our scenario.

\section{Conclusion}\label{sec::sum}

In this article we discussed a novel variant of freeze-out that we call conversion-driven freeze-out. 
In this scenario DM interacts only very weakly with the SM rendering its self-annihilation 
negligible during freeze-out. Nevertheless, a sufficient dilution and cooling of DM can be driven
by a mediator with strong thermal contact to the SM in combination with  conversion processes within the dark sector. 
Typically, the latter require much smaller couplings
to be efficient than annihilation processes during chemical decoupling due to the appearance of light SM 
particles in the initial state with non-Boltzmann 
suppressed number densities. However, for sufficiently small coupling they can initiate the chemical
decoupling and hence govern the relic density. This scenario provides a distinct phenomenology.

In general, the relic density computation in this scenario 
requires the numerical solution of the full coupled set of Boltzmann equations. 
We discuss a simple $t$-channel model with a scalar top or bottom 
partner as a mediator. Both choices provide a cosmologically valid parameter space
in the conversion-driven regime with couplings in the range between $10^{-7}$ and $10^{-3}$ -- out of reach of indirect or direct detection searches in the foreseeable future. 
However, due to a sizeable production cross section at colliders and macroscopic decay length of the mediator,
the scenario can be probed by long-lived particle searches at the LHC\@. While the former
model predicts detector-stable mediators the latter predicts mediator decay length of the order of
the tracker size providing disappearing $R$-hadron tracks and/or displaced $b$-jets.

\section*{Acknowledgements}

J.H.~acknowledges support from the F.R.S.-FNRS, of which he is a postdoctoral researcher as well as
support by the German Academic Exchange Service (DAAD) through the travel grant ``Kongressreisen 2018''.
M.H.~acknowledges supported by the ERC Starting Grant `NewAve' (638528) as well as by the 
Deutsche Forschungsgemeinschaft (DFG, German Research Foundation) under Germany's 
Excellence Strategy (EXC 2121) `Quantum Universe' (390833306).

\bibliographystyle{JHEP}
\bibliography{bibliography}

%
%

\end{document}